\documentclass[journal]{IEEEtran}
\usepackage[utf8]{inputenc}
\usepackage[english]{babel}
\usepackage{booktabs}
\usepackage{multirow}
\usepackage{amssymb}
\usepackage[table,xcdraw]{xcolor}
\usepackage{cite}
\usepackage{float}
\usepackage[pdftex]{graphicx}
\DeclareGraphicsExtensions{.pdf,.png}
\usepackage{subcaption}
\usepackage{xurl}
\UseRawInputEncoding
\usepackage[normalem]{ulem}
 
\usepackage{balance}
\usepackage{algorithm}
\usepackage{algorithmic}
\usepackage[cmex10]{amsmath}
\begin{document}

\title{Integrating HAPS, LEO, and Terrestrial Networks: A Cost-Performance Study for IoT Connectivity
\author{Jean~Michel~de~Souza~Sant'Ana, Felipe~Augusto~Tondo, Nurul~Huda~Mahmood, Aamir~Mahmood.
\thanks{J. M. S. Sant'Ana and N. H. Mahmood are with the Centre for Wireless Communications (CWC), University of Oulu, Finland (\{Jean.DeSouzaSantana, NurulHuda.Mahmood\}@oulu.fi).}
\thanks{F. A. Tondo is with the Federal University of Santa Catarina (UFSC), Florianópolis, SC, Brazil. (felipe.tondo@posgrad.ufsc.br).}
\thanks{A. Mahmood is with the Department of Computer and Electrical Engineering, Mid Sweden University, 851 70 Sundsvall, Sweden  (aamir.mahmood@miun.se).}
\thanks{This research was partially supported in Finland and Sweden by the European Union through the Interreg Aurora project ENSURE-6G (Grant Number: 20361812); in Finland by the Research Council of Finland through the project 6G Flagship (Grant Number: 369116); and in Brazil by the PIPD CAPES-UFSC (88887.103181/2025-00).}}}

\maketitle

\begin{abstract}
This work evaluates the potential of High-Altitude Platform Stations (HAPS) and Low Earth Orbit (LEO) satellites as alternative or complementary systems to enhance Internet of Things (IoT) connectivity. We first analyze the transmission erasure probability under different connectivity configurations, including only HAPS or LEO satellites, as well as hybrid architectures that integrate both aerial/spatial and terrestrial infrastructures. To make the analysis more realistic, we considered movement of LEO satellites regarding a fixed region, elevation angle between gateway and devices, and different fading models for terrestrial and non-terrestrial communication. We also analyze LR-FHSS (Long-Range Frequency Hopping Spread Spectrum) random access uplink technology as a potential use case for IoT connectivity, showing the scalability impact of the scenarios.
The simulation results demonstrate that HAPS can effectively complement sparse terrestrial networks and improve the performance of satellite-based systems in specific scenarios. Furthermore, considering the deployment and operational costs, respectively, CAPEX and OPEX, the economic analysis reveals that although HAPS exhibits higher costs, these remain within a comparable order of magnitude to LEO and terrestrial deployments. In addition, specific use cases, such as natural disasters, transform HAPS into a competitive technology for conventional infrastructures.

\end{abstract}
\begin{IEEEkeywords}
Internet of Things (IoT), Non-terrestrial networks (NTNs), High-Altitude Platform Stations (HAPS), Low Earth Orbit (LEO) satellite.
\end{IEEEkeywords}

\section{Introduction}

Expanding Internet of Things (IoT) connectivity to remote and underserved regions remains a major challenge to achieve global digital inclusion~\cite{Chaoub:WC:2022}. Areas lacking basic network infrastructure, due to geographic, economic, or political constraints, require alternative connectivity solutions that can establish connectivity where none exists or enhance and complement existing systems. Typical applications include herd and wildlife tracking~\cite{Escobar:Sensors:2025}, forest fire detection~\cite{Chan:Access:2024}, agricultural and aquaculture monitoring, offshore ship tracking~\cite{Ullah:ComMag:2022}, and cargo monitoring~\cite{Aslam:WFIoT:2023}.

As a potential solution, Low Earth Orbit (LEO) satellites offer global coverage and scalability~\cite{Fraire.CM.22}. Their reduced costs and recent technological advancements make them an attractive option for extending IoT connectivity. They provide wide coverage per satellite, and their non-stationary orbits enable communication with remote regions. However, a large constellation is required to ensure continuous service and, in most cases, devices must coordinate with the satellites to determine when transmission opportunities occur~\cite{Qi:IoTMag:2024}. Furthermore, the link budget can restrict device operation, especially in indoor environments or areas without line-of-sight, where higher antenna gains, transmission power, or specific configurations may be needed to maintain reliable communication. Furthermore, since LEO satellites simultaneously cover large portions of the Earth's surface due to their high orbital altitude, they can lead to interference challenges in massive IoT scenarios~\cite{Le:CST:2025}.

Another alternative for non-terrestrial networks (NTN)  is the High Altitude Platform (HAPS) mounted network, which is a quasi-stationary aerial system operating in the stratosphere, typically between 8 and 50 km of altitude~\cite{Kurt:2021:CST}. Positioned above weather disturbances, yet much closer to the ground than satellites, HAPS can provide direct communication links to IoT devices with a more favorable link budget and lower round trip time (RTT) latencies. These characteristics allow for intermittent or regional coverage, eliminate the need for gateway synchronization, and offer greater flexibility compared to terrestrial base stations (BS). Recent developments, such as SoftBank's Sunglider~\cite{sunglider:news:2025} and Aalto’s Zephyr~\cite{zephyr:news:2025}, have demonstrated the feasibility of multi-week autonomous solar-powered operation, with both companies planning to begin commercial services in 2026. However, the technology remains at an early stage of maturity, with limited large-scale deployments and cost structures that are still difficult to estimate. In addition, multiple platforms may be required to provide coverage in very sparse or geographically dispersed regions.

In this work, we conduct a comparative evaluation of HAPS and LEO satellite systems for IoT connectivity in remote regions. We also analyze their performance when integrated with existing terrestrial infrastructure. Furthermore, both solutions are compared to terrestrial networks in sparse and moderately deployed gateway scenarios. The main contributions of this work are as follows.
\begin{itemize}
\item A simulation framework for comparing HAPS, LEO satellites, and terrestrial networks in remote areas, each with its own spatial deployment and channel model.
\item An analysis of the success probability for IoT devices communicating directly with HAPS and LEO satellites.
\item An evaluation of hybrid scenarios in which a sparse terrestrial network is complemented by either a HAPS or a LEO satellite.
\item A cost assessment comparing the deployment of HAPS and the utilization of commercial LEO constellation services.
\end{itemize}

The remainder of this paper is organized as follows.
In Section~\ref{sec:stations}, we present the state of the art literature on HAPS and LEO satellite as solutions for remote IoT connectivity. In Section~\ref{sec:model}, we present all the aspects and considerations of the system model used in this work, while in Section~\ref{sec:results}, we analyze and discuss the simulations and numerical results. Finally, we aggregate our thoughts and conclusions in Section~\ref{sec:conc}.

\section{Literature on HAPS and LEO satellites for IoT} \label{sec:stations}

HAPS are quasi-stationary aerial communication platforms that operate in the stratosphere, typically between 8 and 50 km in altitude. They provide wide-area wireless coverage by hosting cellular BS as transmitters or relays, the first of which is particularly interesting for direct-to-device communication. Moreover, HAPS platforms are either aerostatic (e.g., balloons) or aerodynamic (e.g., solar-powered fixed-wing aircraft). Real-world trials and commercial initiatives include SoftBank (previously by HAPSMobile)~\cite{sunglider:news:2025}, Aalto (previously by Airbus)~\cite{zephyr:news:2025} and  World Mobile (previously by Stratospheric Platforms Limited)~\cite{worldmobile:news:2025}, both of which have demonstrated long-endurance flights and high-bandwidth communications. For example, SoftBank’s Sunglider achieved a multi-week stratospheric flight while supporting LTE connectivity. World Mobile has demonstrated direct-to-device 5G capability using a hydrogen-powered platform. Aalto for example holds a record of 67 days flight with its Zephyr platform. These platforms are designed for long missions (weeks to months) and integrate with terrestrial and satellite networks as part of a multi-layered NTN architecture. In terms of deployment, HAPS have been studied and proposed for coverage restoration during disasters (e.g., Germany floods, Japan earthquake), rural broadband in regions like Hokkaido (Japan), and offshore applications including cruise ship coverage and remote oil platforms~\cite{HAPS:2025:WhitePaper}.

The literature on HAPS systems has grown significantly, particularly in the context of 6G and NTN integration. The authors in ~\cite{Abbasi:2024:WC} provide a comprehensive overview of HAPS as a key enabler of vertical heterogeneous networks, emphasizing its ability to provide high-capacity, low-latency coverage over large areas. The paper outlines how HAPS can serve in access and backhaul roles, interacting with both terrestrial and satellite layers, and identifies use cases such as emergency response, rural broadband, and intelligent transportation. The work in~\cite{Lou:2023:WC} focuses on the architectural roles of HAPS in the NTN nexus, presenting quantitative analyzes of routing latency, coverage probability, and energy efficiency in HAPS-assisted ad hoc and cell-free architectures. Their results show significant performance improvements when HAPS are used as relays or as coordination centers in UAV networks.
The HAPS Alliance has released a detailed white paper comparing HAPS, LEO, and GEO satellite capabilities in terms of link capacity, latency, beam coverage, and integration models. It highlights that HAPS can offer 5–50 Gbps throughput at low latency ($<10$ ms) and serve over 100 active beams, enabling efficient support for large-scale IoT and mobile broadband deployments~\cite{HAPS:2025:WhitePaper}.
Kurt et al.~\cite{Kurt:2021:CST} provide a historical and technical survey of HAPS, detailing its evolution from concept to recent trials and its potential role as a super macro base station (SMBS). They analyze the implications for spectrum management, backhaul design, and large-scale deployment feasibility, including challenges related to outage probability and energy sourcing. Moreover, the work in~\cite{Svistunov:arxiv:2025} also mentions the use of HAPS as flying BS as a solution for IoT applications in remote regions.
Several technical works in IoT scenarios focus on the role of HAPS as a relay to a backbone, either from terrestrial or UAV-mounted gateways~\cite{Jia:IoTJ:2023,Andreadis:ICTDM:2023}. Works regarding HAPS as sole BSs are not as common. We can highlight~\cite{Halit:IoT:2022}, where the authors assume a moving HAPS hovering over a certain area. They propose a wake-up mechanism to the LoRaWAN protocol such that devices transmit whenever the HAPS is in coverage and sleep otherwise. Moreover, the authors in~\cite{Giambene:SoftCOM:2022, Almarhabi:ICIAS:2021} evaluate the performance of LoRa network considering a HAPS-mounted gateway, taking into account several aspects such as altitude, transmission power, and collisions between transmissions.
However, important gaps persist in the literature. To the best of our knowledge, only a few studies address HAPS for direct IoT communication, and none provide comparative analyzes against LEO satellite and terrestrial IoT coverage. Furthermore, this is the first work to evaluate coverage in combined architectures while also addressing a cost evaluation on top of it.

Low Earth Orbit (LEO) satellites are the most prominent component of modern NTNs in the context of IoT, offering services from altitudes ranging between 500 and 1,500 km~\cite{EricssonReport:2023}. Their close proximity to Earth enables round-trip latencies in the 25 to 50 ms range, making them suitable for a diverse range of IoT applications~\cite{Fraire.CM.22}. This positions LEO satellites as a viable alternative to traditional terrestrial networks in remote and underserved areas, where other technologies are impractical due to higher costs.
Different from HAPS, there is an extensive literature on LEO satellites for IoT scenarios, as well as several private initiatives with commercial products, showing it as a much more mature alternative. An example is Lacuna Space, which is expanding its satellite IoT connectivity with a constellation of sixteen LEOs under the agreement with Spire Global Data and Analytics~\cite{lacuna2023spire}. Despite advances in satellite communications, several issues remain unresolved or require further investigation for massive IoT data collection. For example, due to the satellite speed, the Doppler phenomenon may work as a barrier, which compromises Direct-to-Satellite (DtS) system performance. In~\cite{Asad.OJOCS.25}, the authors investigate and propose a novel Doppler solution by broadcasting a short Doppler beacon that allows IoT devices to estimate and compensate for the Doppler shift before initiating the uplink transmission phase. Although the results confirm the feasibility of mitigating Doppler effects, several challenges remain with regard to key parameters, such as optimal beacon design. Following the mobility perspective, temporal synchronization between IoT devices and satellites is also considered a major challenge in DtS-IoT networks~\cite{Liang.24.IEEE_N}. The authors in~\cite{Ortigueira.24.SJ} proposed new algorithms that allow IoT devices to predict satellite visibility windows without relying on Internet connectivity, thus facilitating node-to-satellite synchronization. Based on the fact that IoT devices can be aware of satellite trajectories, the authors in~\cite{Tondo:TAES:2025} introduced advanced strategies for DtS-IoT using power domain Non-Orthogonal Multiple Access (NOMA) in the uplink, with fixed or adaptive transmit power. In addition, companies such as Lacuna Space employ predictive techniques to mitigate synchronization issues in various remote scenarios, including agriculture, environmental monitoring, utilities and metering, and maritime connectivity~\cite{lacuna2025application}. The operational concept assumes that satellites broadcast beacons or control messages to synchronize device timing during low-power modes~\cite{LoRa_white_paper.25}. Another important aspect is the design of satellite constellations. The architecture of LEO systems is inherently dependent on large-scale constellations composed of hundreds to thousands of satellites in coordinated orbits~\cite{Qu:Access:2017,Capez:TAES:2022}.

From an economic perspective, studies have been conducted to investigate the viability of using LEO satellites as a potential solution to connect remote areas. The authors in~\cite{Osoro.ACS.21} conducted a technical and economic evaluation of these three constellations, estimating the cost of a single satellite at approximately US\$ 0.6 million, US\$ 5.6 million, and US\$ 3 million for Starlink, OneWeb, and Kuiper, respectively. It is also reasonable to consider that other technologies could be integrated into this scenario to provide better communication conditions. Moreover, authors in~\cite{Toka:2024:CM} perform a total cost-of-ownership (TCO) analysis for three non-terrestrial connectivity technologies: LEO satellite constellations, HAPS, and UAVs. The paper highlights that although HAPS has higher upfront costs, its longer operational lifetime (20 years vs 5 years for LEO satellites) helps narrow the annualized cost gap. However, LEO satellites still benefit from being a more mature technology and show smaller overall costs.

\begin{figure*}[!t]
    \centering
    \begin{subfigure}{0.24\textwidth}
        \includegraphics[width=\linewidth]{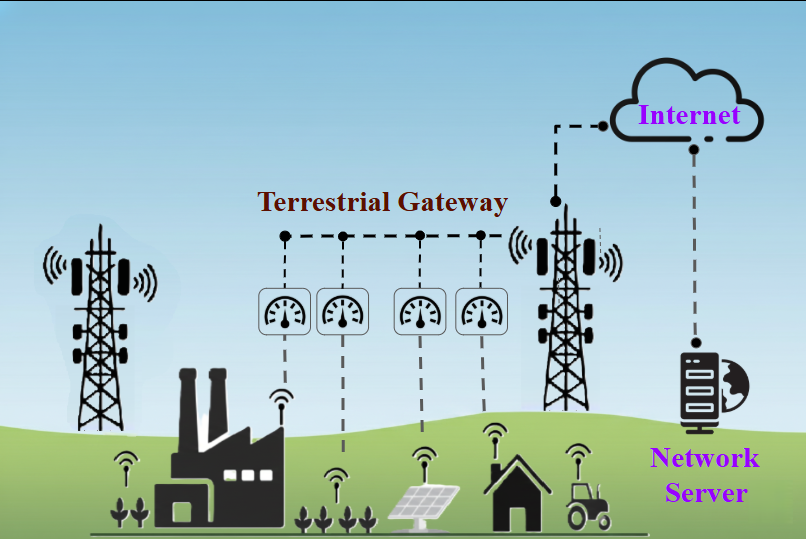}
        \caption{Terrestrial Network (TN).}
    \end{subfigure}
    \begin{subfigure}{0.24\textwidth}
        \includegraphics[width=\linewidth]{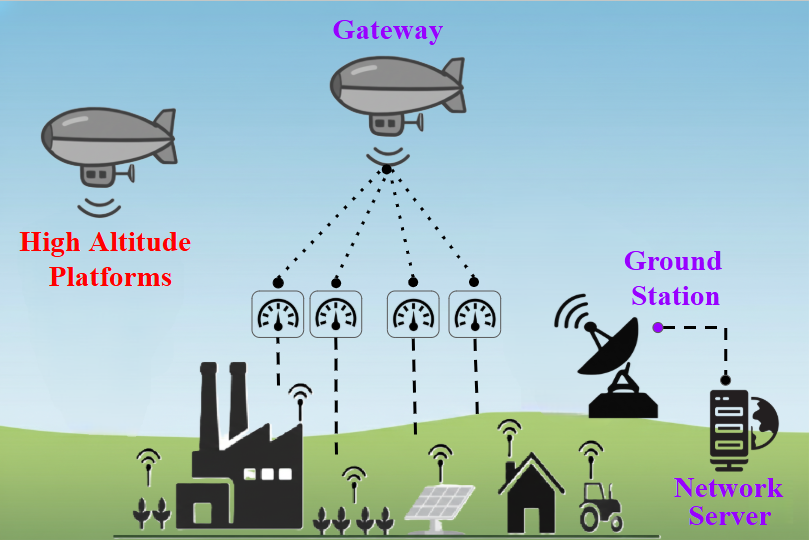}
        \caption{HAPS-based NTN.}
    \end{subfigure}
    \begin{subfigure}{0.24\textwidth}
        \includegraphics[width=\linewidth]{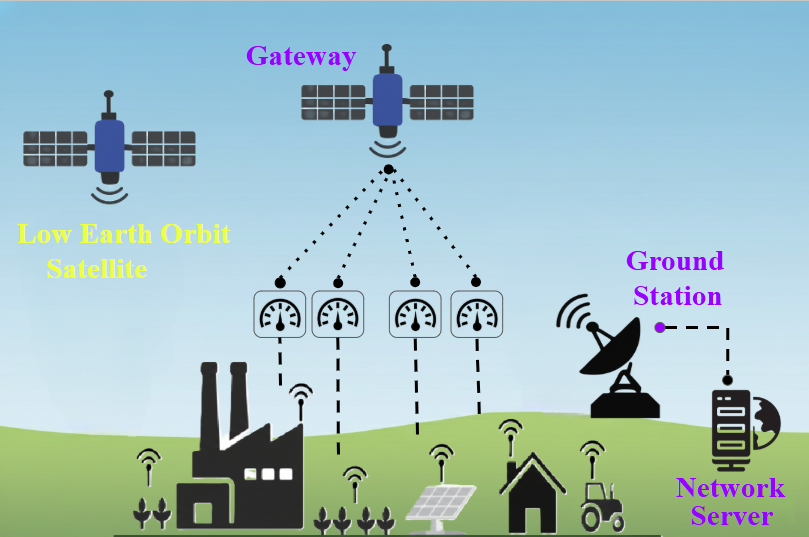}
        \caption{Satellite-based NTN.}
    \end{subfigure}
    \begin{subfigure}{0.25\textwidth}
        \includegraphics[width=\linewidth]{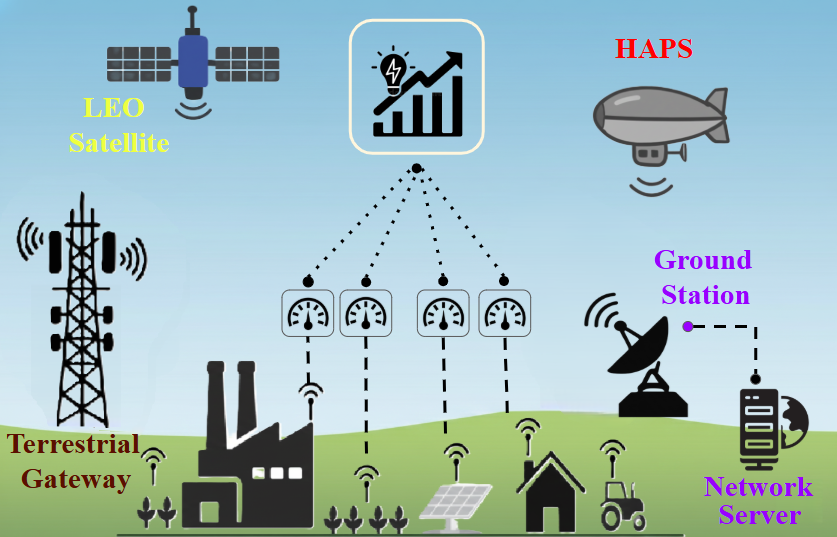}
        \caption{All architectures.}
    \end{subfigure}
    \caption{The illustration of the proposed architectures comprises TNs and NTNs networks, as well as the Ground Station and the Network Server components.}
    \label{fig:system_model}
\end{figure*}

\section{System model} \label{sec:model}

In this section, we introduce the system model and the performance metrics to compare the suitability of HAPS and LEO satellites for remote IoT connectivity. \figurename~\ref{fig:system_model} illustrates an example of several application scenarios, with IoT devices transmitting to terrestrial gateways, HAPS, or LEO satellites. Further, they would relay the information to a ground station backbone or directly to a network server. Note that, for the sake of common notation, we show angles, such as elevation angle and orbit inclination, in degrees. However, all equations are set in radians.

\subsection{Spatial Deployment Model}
First, we start by modeling the distribution of devices and BSs.  We assume that the $N$ devices are distributed at random in a circular area of radius $R$.

The HAPS is considered stationary on top of the center of the circular region, at altitude $h_{\text{H}}$. The distance from a given device $i$ at position $(x_i,y_i)$ to the HAPS is
\begin{equation}
    d_{\text{H}} = \sqrt{x_i^2 + y_i^2 + h_{\text{H}}^2},
\end{equation}
assuming that the devices are at ground level $(z_i=0)$. Moreover, the elevation angle between the device and HAPS is given in radians by
\begin{equation}
\alpha_{\text{H}} = \arcsin \!\left(
\frac{\,h_{\text{H}} R_{\text{e}} - (x_i^2+y_i^2)}
{\sqrt{\bigl(x_i^2+y_i^2+R_{\text{e}}^2\bigr)\,\bigl(x_i^2+y_i^2+h_{\text{H}}^2\bigr)}}
\right),
\end{equation}
where $R_{\text{e}} \approx 6.3781\times10^6$ m is the mean equatorial radius of Earth.

For the satellite, we assume a large enough constellation so that there is always one and only one satellite in coverage, with a minimum elevation angle $\alpha_0$. We also assume a fixed altitude $h_{\text{S}}$, orbit inclination of $i_o=60$ degrees and circular orbit. The period in which one satellite is in coverage is called a lap. The duration of each lap $t_c$ is modeled as a random variable, with its PDF given as~\cite{Seyedi:CL:2012}
\begin{equation}
f_T(t_c) =
\begin{cases}
\displaystyle
\frac{\omega \cos\gamma_0 \tan(\omega t_c)}{\gamma_0 \sqrt{\cos^2(\omega t_c) - \cos^2\gamma_0}}, & 0 < t_c \leq T_m, \\[1.5ex]
0, & \text{elsewhere,}
\end{cases}
\end{equation}
where $\omega=(\omega_s - \omega_e\cos(i_o))/2$, with $\omega_s = \sqrt{\mu_e / r_S^3}$ is the angular velocity of the satellite, $\mu_e = 3.986004418\times10^{14}$ m$^3$/s$^2$ is Earth's gravitational parameter, $r_S = h_S + R_{\text{e}}$ is the satellite altitude from central Earth. Moreover, $w_e = 2\pi/T_s$ is Earth's rotation angular velocity, and $T_s= 86164.1$~s is the sidereal day time. We also have
\begin{equation}
\gamma_0 = \arccos\left( \frac{R_{\text{e}}}{R_{\text{e}} + h_S} \cos\alpha_0 \right) - \alpha_0,
\end{equation}
as Earth's central angle corresponding to the minimum elevation angle $\alpha_0$, where $h_S$ is the satellite altitude. Finally, we have that $T_m = \max(t_c) = \gamma_0 / \omega$ as the maximum coverage time per satellite lap.
With the duration of a lap, we can calculate the elevation angle relative to the origin point of the circular region as a function of time during the given lap $t \in (0,t_c)$ as
\begin{equation}
\alpha_{S_0}(t) = \arctan\!\left(\frac{r_S\cos\!\left(
\left| \frac{\omega t}{2} - \omega t_c \right|\!+\!\gamma_0\!-\!\omega t_c\right)\!-\!R_{\text{e}}}{r_S \sin\!\left(\left| \frac{\omega t}{2} - \omega t_c \right| + \gamma_0 - \omega t_c \right) } \right).
\end{equation}
Note that all variables that depend on $\alpha_{S_0}$ are time-dependent, but we decided to omit them from now on for the sake of clarity.
To calculate the performance metrics of a given device, we first calculate the satellite position vector as
\begin{align}
x_c &= h_S \cos (\alpha_{S_0}\cos\psi),\\
y_c &= h_S \cos(\alpha_{S_0}\sin\psi),\\
z_c &= h_S \sin(\alpha_{S_0}),
\end{align}
where $\psi \! \sim \! U[0,360)$ is the azimuth angle at the central point, which is different for each satellite lap.
Thus, by the Cartesian-to-spherical relation, the elevation angle between the satellite and a given device $i$ at position $(x_i,y_i)$,
\begin{equation}
\alpha_S = 
\operatorname{arctan2}\left(
\frac{z_c}{\sqrt{x_c-x_i)^2 + (y_c-y_i)^2}}
\right).
\end{equation}
Finally, the distance between the satellite and the given device is
\begin{equation}
d_S = \sqrt{(x_c-x_i)^2 + (y_c-y_i)^2 + z_c^2}.
\end{equation}

\begin{figure}[!t]
    \centering
    \includegraphics[width=.9\linewidth]{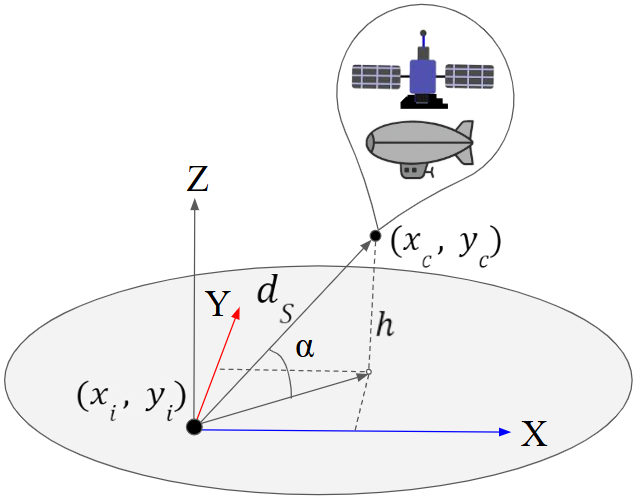}
    \caption{The illustration of the NTN network geometry, which is composed by the device coordinates $(x_i,y_i, z_i)$, elevation angle $\alpha$, altitude $h$ and the distance $d_S$. }
    \label{fig:geometry}
\end{figure}

\figurename~\ref{fig:geometry} depicts the basic geometry of the system model. Note that although shown as a plane, we take into account the effect of a spherical earth when calculating distances and angles.

For the terrestrial network, we assume that $M$ ground BSs are deployed in the circular area at random, such that each BS is separated by a minimum of $s$ km. The separation is to avoid a concentration of BSs in one unique region, which affects the simulation convergence. In addition, we choose $s$ small enough so that $s/2$ represents a reasonable distance for a communication to occur, so that we do not create artificial regions or coverage gaps where communication would fail. Finally, we also consider an annular guard region, with inner radius of $R$ and outer radius of $R+R_g$, where we deploy $M_g$ BSs, so that devices on the border of the circular region do not suffer from the border effect. The number of BSs in the guard region $M_g$ can be defined as
\begin{equation}
    M_g = \operatorname{round}\!\left( M \left(\frac{R_g}{R}\right)\!\left(2+\frac{R_g}{R}\right) \right),
\end{equation}
so that we maintain the same density of BSs in both regions. For simulation tractability, we round it to the nearest integer. Given a device $i$ and a terrestrial BS $j$, at positions $(x_i, y_i)$ and $(x_j, y_j)$, the distance between them is given by
\begin{equation}
    d_{T_j} = \sqrt{(x_i-x_j)^2 + (y_i-y_j)^2}.
\end{equation}

\subsection{Channel model}

We define two different channel models, one for the HAPS and LEO satellite, and another for the terrestrial network. For both cases, we assume that all devices use the same transmit power $p_t$ and carrier frequency $f_c$. Thus, the received power of a single transmission at the gateway from device $i$ is given as
\begin{align}
    p_{rx}^{i} = p_t~g(d_i)~|r|^2,
\end{align}
where $g (d_i)$ is the distance-dependent path loss and $r$ models the fading and other effects, both of which depend on whether the communication is through a terrestrial network or HAPS and LEO satellite.

For the terrestrial network, we model the path loss with an empirical log-normal model from~\cite{Petajajarvi:ITST:2015} as
\begin{equation}
g_T = B + 10\eta\log\left(\frac{d_T}{d_0} \right),
\end{equation}
where $d_T$ is the distance from the device to the BS, $d_0$ is the 1~km reference distance, $B$ is the pathloss at $d_0$, and $\eta$ is the path loss exponent. Moreover, we model a shadow fading $r_T$ as a Gaussian with standard deviation $\sigma_\text{SF}$ in dB.

For communication between devices and the HAPS and LEO satellite, we consider the free space path loss given as
\begin{align}
    g(d) = G_tG_r \left ( \frac{\lambda}{4\pi d} \right )^2, \label{eqn:pathloss}
\end{align}
where $G_t$ and $G_r$ are the antenna gains of the transmitter and receiver, respectively, and $\lambda=c/f$ is the signal wavelength with $c=3\times10^8$ m/s as the speed of light. Note that the receiver antenna gain depends on whether we consider HAPS ($G_{R_H}$) or LEO satellite ($G_{R_S}$).

We model channel fading for HAPS and LEO satellite communication $r_H$ as \cite{Abdi:TWC:2003}, a shadowed Rice model with LOS amplitude characterized by the Nakagami distribution. This model is relevant to our scenario, as it accounts for the elevation angle's impact on fading and shadowing. Moreover, this model can be approximated by a gamma distribution~\cite{Talgat:TAES:2024}, which results in easier mathematical tractability. The PDF of $|r_H|^2$ is~\cite{Talgat:TAES:2024}
\begin{align}
    f_{|r_H|^2}(h) = r_H^{k-1} \frac{e^{-r_H/\theta}}{\theta^k \Gamma(k)},
\end{align}
where $\Gamma(x)$ is the Gamma function, $k = \frac{m(2b_0 + \Omega)^2}{4mb_0^2+4mb_0\Omega + \Omega^2}$ and $\theta = \frac{4mb_0^2 + 4mb_0\Omega + \Omega^2}{m(2b_0+\Omega)}$ are the shape and scale parameters, and $m$, $b_0$ and $\Omega$ are empirical expression parameters derived as a function of the elevation angle $\alpha$, given as
\begin{align}
    b_0(\alpha) &= -4.7943 \times 10^{-8}\alpha^3 + 5.5784 \times 10^{-6}\alpha^2 \nonumber \\
    &\quad - 2.1344 \times 10^{-4}\alpha + 3.2710 \times 10^{-2} \nonumber \\
    m(\alpha) &= 6.3739 \times 10^{-5}\alpha^3 + 5.8533 \times 10^{-4}\alpha^2 \nonumber \\
    &\quad- 1.5973 \times 10^{-1}\alpha + 3.5156 \nonumber \\
    \Omega(\alpha) &= 1.4428 \times 10^{-5} \alpha^3 - 2.3798 \times 10^{-3} \alpha^2 \nonumber \\
    &\quad + 1.2702 \times 10^{-1}\alpha - 1.4864.
\end{align}

\subsection{Performance metrics}

In this work, we assume LoRaWAN LR-FHSS technology to evaluate the performance of the scenarios. LR-FHSS is a frequency-hopping uplink scheme in LoRaWAN that uses GMSK modulation over 488~Hz physical channels. In our case, we consider the European ISM carrier frequency of 868~MHz, and each transmission hops over 35 such channels following a pseudorandom sequence derived from the header seed. Each packet contains three header copies, each lasting 233.472~ms, sent on different physical channels. The payload is convolutionally coded at a coding rate of $\textsf{CR}=1/3$ and split into fragments of 102.4~ms. The number of fragments is
\begin{equation}
    f = \left\lceil \frac{b+3}{6\,\textsf{CR}} \right\rceil ,
\end{equation}
which for $b=10$ bytes and $\textsf{CR}=1/3$ yields $f=7$. Devices use unslotted ALOHA, randomly selecting a seed and hopping to a new physical channel for every header and fragment. A receiver that decodes at least one header can reconstruct the hop sequence. The packet is successfully decoded if the header and at least one third of the fragments (rounded up) are received.

We use two metrics to evaluate performance. The first is the erasure condition, which is the probability that the received signal power is below a given technology threshold $\gamma$. This is the most interesting performance metric for this work, as it directly shows the impact of the topology of the scenarios, regardless of the number of devices. This probability can be written as
\begin{equation}
    C_1 = \text{P}\left[p^i_{rx} < \gamma\right].
\end{equation}
For the LR-FHSS scenario, we consider that each header or fragment perceives a different realization of $p^i_{rx}$. However, they will be correlated because of the elevation angle (in the case of NTN) and the distance between the device and the gateway. Although LR-FHSS has coding and replication schemes, we believe it is most important to analyze the erasure probability of single headers and fragments, as this value could easily be translated to any other technology, with the proper $\gamma$.

The second metric is the packet success probability. For each packet, we assume it to have been successfully decoded if at least one header and $\textsf{CR} = 1/3$ of the payload fragments have been decoded. Headers and fragments are susceptible to erasure as in the first metric, and collisions, in case other headers or fragments transmissions occur at the same in the same physical channel. We analyze the LR-FHSS transmission success probability using an LR-FHSS network simulator~\cite{Santana:EuCNC:2024}. Here, we assume that all collisions are destructive, i.e., no capture effect\footnote{The simulator was adapted to account for the effects of the scenarios in this work, and the repository with the adapted version of the simulator will be openly published and attached to this manuscript once accepted}. The same packet can be decoded by multiple BSs, while we assume a centralized network server responsible for dealing with duplicates.

\subsection{Costs analysis}

Here we present the cost models for all scenarios. Inspired by~\cite{Toka:2024:CM}, we consider the \texttt{total cost} by bringing it to the present value, with a discount factor $\sigma$, to take into account the opportunity costs. Thus, we have
\begin{equation}
    \texttt{total cost} = \text{CAPEX} + \sum_{n=1}^{\texttt{years}}\frac{\text{OPEX}}{(1 + \sigma)^{n}},
\end{equation}
where CAPEX represents the deployment costs, OPEX the yearly operational costs and \texttt{years} the time-window horizon in years.

According to~\cite{Toka:2024:CM}, the CAPEX costs for HAPS are \$4M, with an OPEX of \$30,000. In addition, they utilize a discount factor $\sigma=0.05$ and a window of 20 years, both of which we will adopt in this work. For the LEO satellite, we consider a commercial connectivity provider as the reference for satellite-based NTN services. In this context, companies such as Wyld Networks~\cite{WyldNetworks.25}, Myriota~\cite{Myriota.25}, and LacunaSpace~\cite{lacuna2023spire} offer satellite connectivity on a pay-per-message or subscription basis, partnering with LEO satellite operators to deliver data from remote devices to the cloud. According LacunaSpace page, they offer services of around \$24 per device per year, where this value could be smaller for services of more than 5000 devices. Thus, we have a null CAPEX, with an OPEX of \$$24\times N$. Finally, for terrestrial BSs, we account the average yearly tower leasing according to~\cite{SteelInTheAir.25}, which is \$12,600. Thus, we have again a null CAPEX and an OPEX of \$12,600$\times M$.

\section{Numerical Results and Discussions} \label{sec:results}

In this section, we analyze the numerical results from the model proposed in the previous sections. If not explicitly mentioned, we utilize the simulation parameters in Table~\ref{tab:parameters}. We picked scenarios with a single HAPS, a system with always one LEO satellite in coverage, a terrestrial network with 10 and 30 BSs, and a combination of them. With that in mind, we intend to analyze the performance of each system and also evaluate the benefits of using them together. Finally, we understand that with 10 BSs we have a sparsely deployed network, with low coverage and several gaps, while 20 BSs represent a scenario with relatively good coverage.

\begin{table}[]
\centering
\small
\caption{Simulation parameters}
\begin{tabular}{@{}lll@{}}
\toprule
\textbf{Parameter}       & \textbf{Symbol}   & \textbf{Value}    \\ \midrule
Network radius  & $R$                     & 80 km    \\
HAPS altitude  & $h_H$                  & 30 km    \\
Satellite altitude  & $h_S$                 & 750 km   \\
Orbit inclination  & $i_o$                  & 60º      \\
Reference distance  & $d_0$                 & 1 km     \\
Log-normal path loss exponent  & $\eta$     & 2.32     \\
Path loss at the reference distance  & $B$  & 128.96 dB  \\
Shadow fading standard deviation  & $\sigma_{\text{SF}}$ & 7.8 dB \\
Minimum terrestrial BS separation  & $s$                     & 20 km    \\
Guard region radius  & $R_g$                  & 240 km   \\
Transmission power  & $p_t$                  & 14 dBm   \\
Carrier frequency  & $f_c$                  & 868 MHz  \\
HAPS antenna gain  & $G_{R_H}$               & 6 dBi    \\
Satellite antenna gain  & $G_{R_S}$               & 13.5 dBi \\
Terrestrial BS antenna gain  & $G_{R_T}$               & 6 dBi    \\
Payload size  & $b$                     & 10 bytes \\
Receiver sensitivity  & $\gamma$             & -132 dBm \\ 
Average transmission interval  & - & 15 minutes\\
Minimum elevation angle & $\alpha_0$ & 20 degrees \\
\bottomrule
\end{tabular}
\label{tab:parameters}
\end{table}

\subsection{Erasure Probability}

Each figure scenario in this subsection represents data from headers and fragments transmissions for 5000 simulation runs, each one with 1000 devices over a period of 24 hours. Each data point is the average erasure experienced by a device in given position during each simulation.

\begin{figure*}[tb]
    \centering
    \includegraphics[]{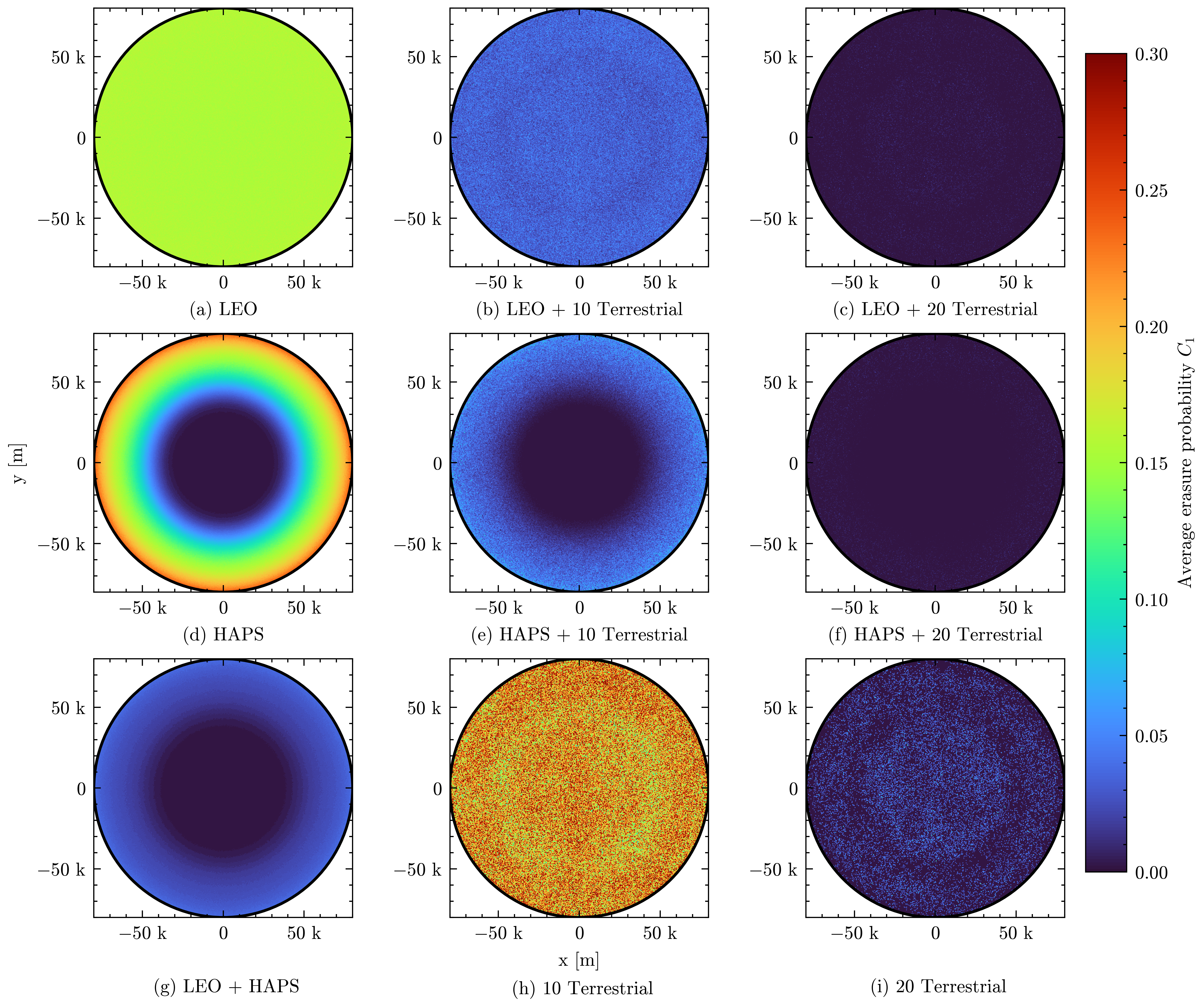}
    \caption{Heat map of the erasure probability over the area where devices are deployed for all proposed scenarios.}
    \label{fig:heatmap_erasure}
\end{figure*}

\figurename~\ref{fig:heatmap_erasure} depicts a heat map of the erasure probability in the circular area where the devices are deployed. Thus, we can identify the probability that a transmission will fail to reach a BS. 
First, we can see in \figurename~3a that the LEO satellite model shows a uniform distributed coverage, which means that the geographical positioning of the device in the area does not affect performance. A similar idea applies to the terrestrial network case in \figurename~3h, however, it presents a higher variance due to the dependency on the distribution of BSs over the area. The HAPS scenario in \figurename~3d  presents devices in the network border with worse performance compared to the ones in the center, as expected. We can see that 10 terrestrial BSs in \figurename~3h present coverage issues, going under 30\% in several points. Moreover, depending on the BS deployment, some areas can be without any coverage. Moreover, while HAPS presents an unfair distribution compared to LEO, we can see that HAPS presents a lower erasure probability in a large portion of the area. In addition, the combination of HAPS with the terrestrial network (\figurename~3e and \figurename~3f)  works in favor of reducing the worse unbalanced performance on the network border. Moreover, it presents a lower erasure probability than LEO satellite for both cases in \figurename~3b and \figurename~3c. Finally, a combination of LEO and HAPS in \figurename~3g shows to outperform the terrestrial case with 20 BSs in \figurename~3i, which presents an interesting idea in which the integration of multiple different non-terrestrial systems might be even more competitive than dense terrestrial network systems in terms of coverage/erasure probabilities.

\begin{figure*}[tb]
    \centering
    \includegraphics[width=\linewidth]{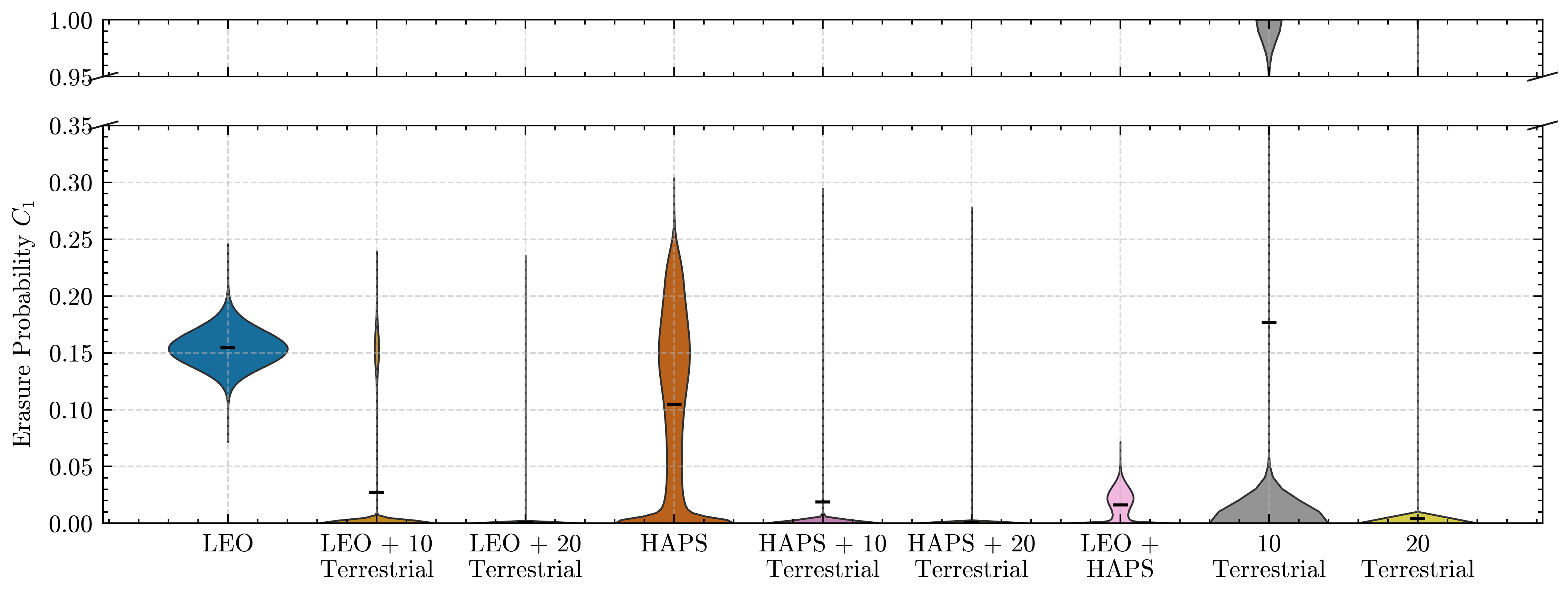}
    \caption{Erasure probability violin graph for all proposed scenarios. Wider graphs represent a higher density of that probability and the white line shows the median.}
    \label{fig:violin_erasure}
\end{figure*}

\figurename~\ref{fig:violin_erasure} shows a violin graph with the distribution of the erasure probability in the aforementioned scenarios. In this figure, the height of each curve represents the distribution of the erasure probability, while the width shows the density of the erasure probability value. Each violin has to be interpreted separately, as the width is accounted for individually. Finally, a black line on them represents the average erasure probability for each case. As seen before, the LEO probability varies less over the area compared to the other cases. In this figure, the effect is represented as a smaller vertical spread. In the HAPS case, we see a higher variation due to the geographical distribution seen before. However, although the terrestrial erasure probability did not change over the area, a high variance is observed even in the scenario with 20 terrestrial BSs. This shows us that the performance of this system is highly dependent on the distribution of the BSs. Even guaranteeing a certain distance between the BSs, the scenario presents cases where devices did not have coverage. On the other hand, the same behavior was not seen in the non-terrestrial scenarios. Also, the terrestrial and non-terrestrial integration scenarios did not present this very high variance, showing that they can help mitigating the deployment dependence of terrestrial networks. Finally, the combination of LEO and HAPS presents a slightly higher erasure probability average than the 20 terrestrial BSs, in exchange of the benefit of guaranteeing coverage fairness over the whole area without the need to take into account BS deployment planning.

\subsection{Success Probability}

\begin{figure*}[tb]
    \centering
    \begin{subfigure}{\columnwidth}
    \includegraphics[width=\linewidth]{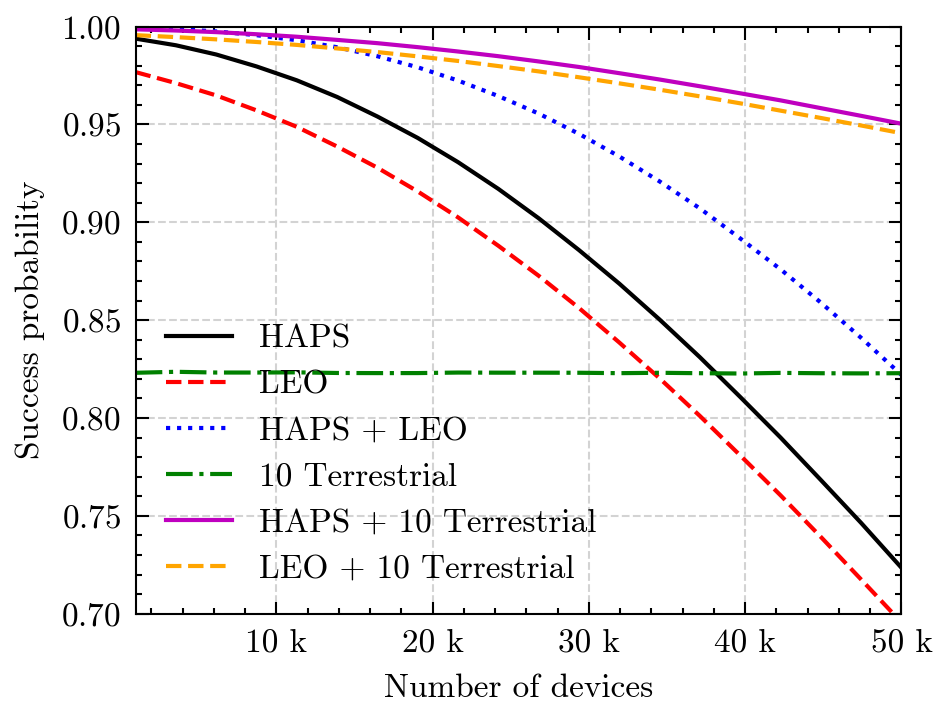}
    \caption{$M=10$}
    \label{fig:success_T10}
    \end{subfigure}
    \begin{subfigure}{\columnwidth}
    \includegraphics[width=\linewidth]{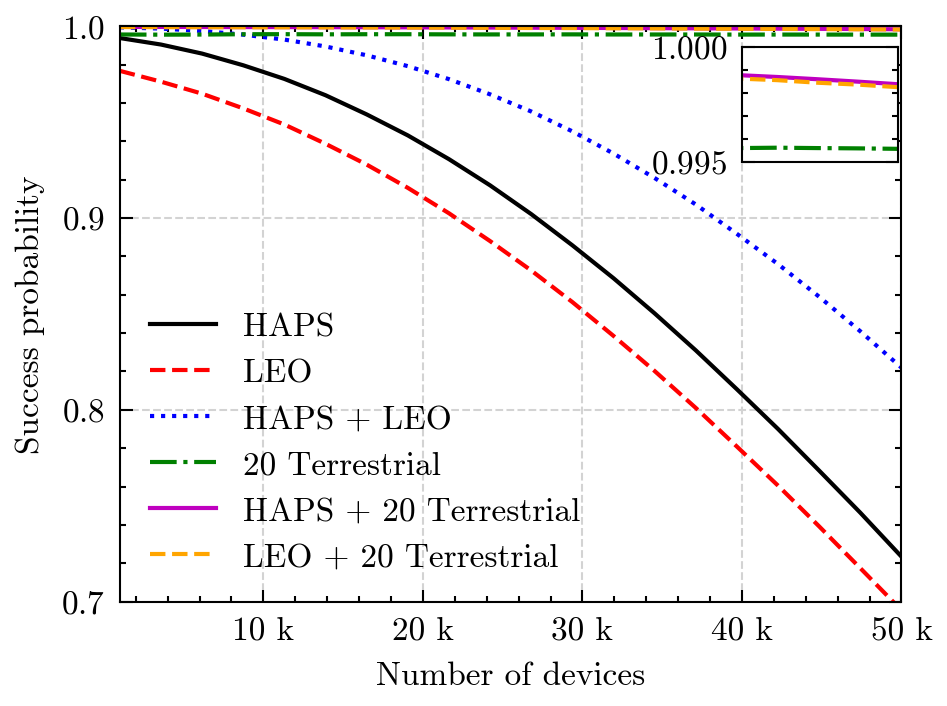}
    \caption{$M=20$}
    \label{fig:success_T30}
    \end{subfigure}
    \caption{Success probability as a function of the number of devices for all proposed scenarios.}
    \label{fig:success_probability}
\end{figure*}

In this subsection, each data point is acquired by averaging the success probability of all devices in each run, and then averaging over 5000 runs with 1 hour duration each.
\figurename~\ref{fig:success_probability} shows the transmission success probability for the proposed scenario. This analysis is particularly interesting when we plan to understand the network scalability in scenarios with a large number of devices active in the system. The first behavior to note in both cases is that in the case of terrestrial networks, both for 10 and 20, the success rate changes little with the number of nodes. Since path loss tends to act stronger on terrestrial communication, it is common that each terrestrial BS has less interference from devices transmitting from far away. The number of devices seen by each BS is smaller, thus there are fewer collisions happening compared to the non-terrestrial scenarios. In~\figurename~\ref{fig:success_T10} we illustrate that the performance of the terrestrial network is very low, due to the high erasure probability seen in the previous section. We see HAPS slightly outperforming LEO, and the combination of LEO and HAPS outperforming the integration of HAPS and LEO with terrestrial BSs until around 10k devices. This tells us that even sparsely deployed terrestrial BSs can help in the scalability of the network by integrating with non-terrestrial options. In \figurename~\ref{fig:success_T30},  the terrestrial network exhibits a very high success probability, proving that it is in fact a scenario under strong coverage conditions. The integration of LEO and HAPS with the terrestrial network still improves the success probability, even if the number of IoT devices increases within the network. This is highlighted by the purple and orange lines at the top of \figurename~\ref{fig:success_T30}.

\subsection{Connectivity Costs Analysis}

In this section, we evaluate the costs considering HAPS, LEO satellite and terrestrial networks. To simplify and provide a more realistic point of view, we consider a commercial connectivity provider as the reference for satellite-based NTN services. Also, due to the lack of commercial services of HAPS, we could not find any data for it as a service. In addition, we did not consider the deployment of a LEO constellation, as it is intended to cover multiple regions due to its non geostationary behavior, and thus the cost would be higher. The cost values adopted in our analysis are derived from publicly available information and indicative pricing from such providers, enabling a practical comparison with the potential costs of deploying a HAPS-based connectivity solutions. The costs of terrestrial networks are related to the leasing costs of communication towers, while we disregard the hardware costs.

\begin{table}[tb]
\centering
\caption{Main cost parameters for terrestrial and non-terrestrial infrastructures.}
\label{tab:costs}
\begin{tabular}{lccc}
\hline
\textbf{Parameter} & \textbf{HAPS} & \textbf{Satellite} & \textbf{Terrestrial} \\ \hline
Ref. & \cite{Toka:2024:CM} & \cite{WyldNetworks.25, Myriota.25, lacuna2023spire} & \cite{SteelInTheAir.25} \\
Model Type & Dedicated platform & Pay-per-device & Tower leasing \\
CAPEX (USD) & \$4M & - & - \\
OPEX (USD/year) & \$30,000 & \$24$ \times N$ & \$12,600$ \times M$ \\
Discount Factor ($\sigma$) & 0.05 & 0.05 & 0.05 \\
Total Time (years) & 20 & 20 & 20 \\

 \hline
\end{tabular}
\end{table}

\begin{figure}
    \centering
    \includegraphics[width=\linewidth]{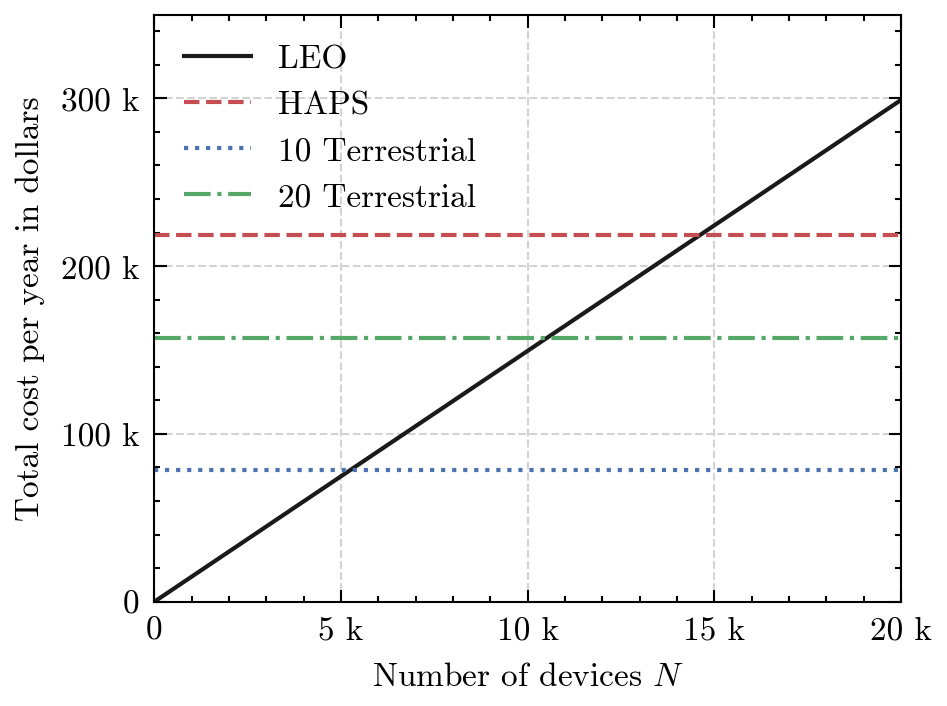}
    \caption{Total cost per year as a function of the number of devices for LEO, HAPS and terrestrial networks.}
    \label{fig:costs}
\end{figure}

The values used in the analysis are shown in~\ref{tab:costs}. We assume a 20 year span time analysis for fairness, since our data assume the HAPS system would last for at least 20 years. Fig.~\ref{fig:costs} presents the cost considering HAPS, LEO and terrestrial solutions as a function of the number of devices, assuming one message per hour over one month. We can see that even though the HAPS costs are higher, they are already competitive with the costs of 20 terrestrial BSs. The LEO solution is shown to be interesting when the number of devices is not too large ($< 10k$), although lower prices can even be negotiable with these providers in case of this massive number of devices. This shows that LEO solutions are interesting in the case that the number of devices is not too large, at a price of slightly smaller performance than HAPS, while being more flexible than a terrestrial network. In addition, terrestrial networks tend to perform better when well deployed, subject to deployment availability, as its costs may increase significantly if there is a need to install non-existing infrastructures. The authors also acknowledge that the simplicity of the cost analysis, especially in terms of scalability, as most likely the LEO service would give a quality of service guaranty, either by improving the infrastructure or increasing the operational bandwidth. On the other hand, the constant values in the owned infrastructure options actually decrease the performance per-device as the number of devices increases, as seen in~\figurename~\ref{fig:success_probability}.

\section{Conclusions} \label{sec:conc}
In this work, we compared the impact of HAPS and LEO satellites as an alternative to IoT connectivity in remote regions with a lack of terrestrial connectivity. We also consider cases where HAPS and LEO could work together with existing terrestrial networks and also a combination of both HAPS and LEO. We analyze the transmission erasure probability behavior and conclude that HAPS can be a good addition to existing sparse terrestrial networks, and also to the LEO satellite system. Moreover, we also conducted an economic analysis of the deployment of an HAPS versus using LEO satellite IoT services, as well as the deployment of terrestrial networks.
The results highlight that HAPS outperforms LEO in terms of success probability and average erasure probability, and that the combination of sparse terrestrial networks with non-terrestrial solutions as a potential enabler for massive connectivity.
Finally, we show that although the costs of HAPS are higher, they were shown to be in a similar magnitude order, and with the advancements of this technology, they can be an promising alternative to IoT connectivity. Future work includes the investigation of smarter diversity techniques to exploit simultaneous reception from multiple receivers, the incorporation of multi-satellite visibility and HAPS mobility into the system model, and the design of transmission strategies that leverage knowledge of HAPS and LEO movement patterns.

\bibliographystyle{IEEEtran}
\bibliography{references}

\end{document}